\documentclass[12pt,preprint,epsf]{aastex}

\def\spose#1{\hbox to 0pt{#1\hss}} 

\def\simlt{\mathrel{\spose{\lower 3pt\hbox{$\mathchar"218$}}      

\raise 2.0pt\hbox{$\mathchar"13C$}}} 

\def\simgt{\mathrel{\spose{\lower 3pt\hbox{$\mathchar"218$}}      

\raise 2.0pt\hbox{$\mathchar"13E$}}}

\def\etal{{\rm et~al.~}} 

\righthead{Exploring the Dark Sector } 

\begin{document}


\title{The Hubble constant and dark energy}

\author{Jeremy Mould} 

\affil{Centre for Astrophysics and Supercomputing, Swinburne University} 

\authoraddr{E-mail: jmould@swin.edu.au}


\begin{abstract}

The Hubble Constant measured from the anisotropy in the cosmic microwave background (CMB) is shown to be independent of small changes from the standard model of the redshift dependence of dark energy. Modifications of the Friedmann equation to include phantom power (w $<$ --1), textures (w = --2/3) and curvature are considered, and constraints on these dark energy contributors from supernova observations are derived. Modified values for the density of matter inferred from cosmic density perturbations and from the CMB under these circumstances are also estimated, as exemplified by 2df and SDSS.

\end{abstract}

\section{Introduction}

Turner (1999) coined the term dark energy to name the power source of the accelerating
Universe (Garnavich \etal 1998; Perlmutter \etal 1999). 

For the first time, there was a plausible, complete accounting of matter and energy in the Universe. Expressed as a fraction of the critical density there are: neutrinos, between 0.3\% and 15\%; stars, 0.5\%; baryons (total), 5\%; matter (total), 40\%; smooth, dark energy, 60\%; adding up to the critical density. This accounting is consistent with the inflationary prediction of a flat Universe and defines three dark-matter problems: Where are the dark baryons? What is the nonbaryonic dark matter? What is the nature of the dark energy? The leading candidate for the (optically) dark baryons is diffuse hot gas; the leading candidates for the nonbaryonic dark matter were slowly moving elementary particles left over from the earliest moments (cold dark matter), such as axions or neutralinos; the leading candidates for the dark energy involve fundamental physics and include a cosmological constant (vacuum energy), a rolling scalar field (quintessence), and light, frustrated topological defects. 

Gooding \etal (1992) considered a Universe in which density fluctuations are produced
in an initally smooth Universe by the ordering dynamics of scalar fields following a symmetry
breaking phase transition at the grand unified scale. Such transitions lead
to the formation of an unstable topological defect known as ``global texture."

Carroll \footnote{ (http://ned.ipac.caltech.edu/level5/Carroll2/Carroll2$\_$1.html)} points out that
for some purposes it is useful to pretend that the $-ka^{-2} R_0^{-2}$ term in the
Friedmann equation represents an effective ``energy density in curvature'', and define $\rho_k$ 
$-(3k / 8\pi GR_0^2) a^{-2}$.

Caldwell (1999 astro-ph 8168) remarks that most observations are consistent with models right up to the w = --1 or cosmological constant limit, and so it is natural to ask what lies on the other side at w $<$ --1. He termed this phantom energy. 

In this paper we outline how a dark energy program will constrain these elements and, in particular,
how they affect the measurement of the Hubble Constant by means of the anisotropy in the cosmic microwave background.
Section 2 extends the Friedmann equation; $\S$3 shows that supernova data are currently
tolerant of small values of $\Omega_1$ and $\Omega_2$; $\S$4 explores the degeneracies
in CMB data; $\S$5 examines how matter density experiments like 2dF (Peacock \etal 2001) 
are affected; $\S$6 broadly explores the parameter space of $\Omega_n$ as it applies
to SN and CMB data. Our conclusions are in the final section.

\section{The Expansion Rate}

An observer confronted with data like that in Figure 1 might respond by fitting a polynomial
to the expansion rate as a function of redshift. But a physical equation already exists,
namely the Friedmann equation.

$$(H/H_0)^2 = \Sigma_{-1}^4 a^{-n} \Omega_n = h^2    \eqno(1)$$

From the point of view of fitting the data the observer might be surprised at the emphasis placed by physics on the higher order coefficients. This was not rectified until the
discovery of dark energy, based on earlier versions of Figure 1 by the High z Supernova
and Supernova Cosmology teams, although the zeroth order coefficient was considered and discarded 
by Einstein.

According to Gooding \etal (1992) the textures source term is

$$S_T = 4\pi G(\rho_T + 3P_T)\tau_*^2 a^2/(1+a)   \eqno(2)$$

where $\tau_*$ is a time constant equal to $(8\pi G\rho_{eq}/3)^{-1/2}$
and the scale factor, a, is taken to be unity at the equality of matter and radiation.
The quantities $\rho_T$ and $P_T$ are the density and pressure of textures respectively.
The time constant is just the age of the Universe at equality.
When a $>>$ 1, S scales like $\rho a$, and textures will contribute to $\Omega_1$.
When a $<<$ 1, S scales like $\rho a^2$, and textures will contribute to $\Omega_2$.

The $\Omega$ coefficients are normalized by the Friedmann equation, so that

$$\Sigma_{n=-1}^4 \Omega_n = 1 \eqno(3)$$

where

$$n = 3 ( 1 + w_n) \eqno(4)$$
specifies the equation of state for matter and radiation components etc.

\section{Fitting the Supernova Observations}


The current supernova data (Conley \etal 2011) have been processed by Ned Wright and are shown in Figure 1.
A value of $\Omega_k$ = 0.05 does not violate the data. GRB data are also shown in Figure 1 (Schaefer 2006).
The data are reproduced in Figure 2.
Assuming $\Omega_{-1}~=~\Omega_2$ = 0, a value of $\Omega_1$ = 0.1 does not violate the data.
The data are reproduced again in Figure 3, where we consider the case w = --4/3.
Assuming $\Omega_{1}~=~\Omega_2$ = 0, 
a value of $\Omega_{-1}$ = 0.1 does not violate the data.
Larger doses of textures, curvature, and phantom power would violate the data.
We revisit these data to place firm constraints in $\S$6.

\section{Fitting the CMB}

Some constraints on $\Omega_{\pm 1}$ and $\Omega_2$ are imposed by the small scale anisotropy of the cosmic microwave background.
Komatsu \etal (2009) deduced --0.0179 $<~\Omega_k~<$ 0.0081 (95\% confidence).

We can derive similar constraints on $\Omega_n$ generally by requiring that
the acoustic scale and shift parameter, R, (Komatsu \etal 2011) are conserved. We also invoke equation (3).
For small values of the textures and curvature contributions, writing $\delta R ~=~ \Sigma \partial R /
\partial \Omega_n \delta \Omega_n$ and a similar expression for the acoustic scale,

$$\Sigma_{n=0}^2f_n\delta\Omega_n =0; ~~~~~~~~~\delta\Omega_3 = 0; ~~~~~~~~~\Sigma_{n=0}^2\delta\Omega_n = 0 \eqno(5)$$
where 
\footnote{Equation (5) is simply derived for $dr_s/d\Omega_3 = 0 ~{\rm and}~ dz_*/d\Omega_3 = 0$, but is also correct
when the exact density dependence of the sound horizon, $r_s$, and CMB redshift, $z_*$, is included.}
$$f_n = \int_0^z(1+z^\prime)^nh^{-3}(z^\prime)dz^\prime \eqno(6)   
$$

The acoustic scale and shift parameter are conserved
when introducing $\Omega_1$ = $\delta$ to the WMAP model provided
$\delta \Omega_n ~=~ c_n \delta$
where $c_n$ are coefficients of order unity and $c_1$ = 1, $c_3$ = 0,
$c_0=(f_1-f_2)/(f_2-f_0)$, and $c_2 = -(f_1-f_0)/(f_2-f_0)$  are computed in a simple numerical integration.
Values are given in Table 1.
For example, if $\Omega_1$ = 0.1, $\Omega_0$ = 0.73 -- 0.08 = 0.65,
and $\Omega_2$ = --0.02.
Similar equations can be written for $\Omega_{-1}$ if we adopt $\Omega_1$ = 0.
The change in the Hubble Constant deduced by WMAP is proportional to $\Sigma f_nc_n$  which is zero.

\section{Density perturbations}

Density perturbations in the Universe evolve as

$$\delta_{grow} \propto H \int_0^a da/\dot{a}^3 \eqno(7)
$$

By requiring changes in $\delta_{grow}$ relative to those detected by 2dF and SDSS
to be zero in response to $\delta \Omega_n$, we can follow the formalism of the previous section to obtain

$$\Sigma_{n=0}^3g_n\delta\Omega_n =0; ~~~~~~\Sigma_{n=0}^3\delta\Omega_n = 0 \eqno(8)$$
where
$$g_n = \int_{0.001}^1 a^{-(n+3)}h^{-5}(a)da   \eqno(9)$$.

Calculating the $g_n$ values numerically (see Table 1), we find that the growth factor is conserved when $\delta\Omega_1$ and $\delta\Omega_2$ are introduced to
the 2dF/SDSS model provided 

$$\delta\Omega_3 = -0.127\delta\Omega_1 -0.384\delta\Omega_2   \eqno(10)$$.

So perturbing the standard model by 0.1 in $\Omega_1$, one would perturb the matter density measurement by only --0.01. And perturbing the standard model by 0.1 in $\Omega_k$, one would perturb the density measurement by --0.04. This would be a significant change.
Similar equations can be written for $\Omega_{-1}$ if we adopt $\Omega_1$ = 0.
\section{Constraining $\Omega_{0,1,2}$ in a dark energy program}
\subsection{Combined SN and CMB constraints}
From WMAP7 we formed the data vector $(l_A,R,z_*)$ and calculated $\chi^2$ for a full grid
of values of $\Omega_n$. For SNe $\chi^2$ was calculated directly from the data shown in
Figure 1.
Marginalising over $\Omega_m$, we can calculate probability in the ($\Omega_{\pm 1}, \Omega_2$) plane given the SN data in Figure 1 and the WMAP 7 year data. The results are in Figures 4 and 5. We confirm what we found in section 4, that the Hubble Constant and the density of matter do not constrain these parameters.
\subsection{The expansion rate at larger redshifts} 
A polynomial approach to dark energy in the Friedmann equation may actually lead to physical insights. 
Because unknown physical processes may be classified by how they scale with 1+z,
they can at least be ranked by our approach. 

Quintessence is beyond the scope of the present work. 

$\Omega_{0,1,2}$ can be measured via experiments\footnote{For example, $\Omega_2$ = 72$\delta_2 h^2$ -- 48$\delta_1 h^2$ -- 0.3, 
where $\delta_z h^2 = h^2(z)-h^2(1)$.
If only one of these parameters is nonzero, e.g. $\Omega_2$, then 
$\Omega_2$ = 7.2$\delta_2 h^2$ --0.17 and it could be measured to 10\% accuracy by a differential expansion rate 
experiment of similar measurement precision.}
 to determine $\delta h^2$ at z = 2 and z = 3, 
where h is the dimensionless expansion rate, h(z). 

Of course, our enthusiasm for polynomials should not obscure the real purpose of a dark energy program which is to 
determine the expansion as a function of redshift and the underlying physics, not simply an analytic form of the Friedmann equation. 

\subsection{Dark Energy Surveys}
Experiments such as the Dark Energy Survey (Frieman \etal 2005) and WiggleZ (Blake \etal 2011) are aimed at determination of the 
equation of state P = $\rho$w. Measurement of $\Omega_1$ and $\Omega_0$ are also within their scope (Komatsu \etal 2009; equation 80).

$$a^{3(1+w_{eff})} = \Omega_\Lambda / (\Omega_0+\Omega_1/a) \eqno(11)$$

For small z and $w_{eff} \approx$ --1, 

$$3w^\prime_{eff}z = \Omega_1/\Omega_0 = 0 ~{\rm for}~ \Omega_1 = 0   \eqno(12)$$

where $w^\prime~=~dw/da$.

Coefficients in the Friedmann equation are related to w by equation (4)
and are identified in Table 1.
\vskip 0.5 truein

\leftline{\bf Table 1: Equation of state components}

\begin{tabbing}

ssssssssssssssss\=sssssssssss\=ssssssssssss\=sssssssssssss\=sssssssssssss\=sssss\kill
$\Omega_n$\>n\>w$_n$\>c$_n$\>f$_n$\>--g$_n$\\
phantom\>--1\>--4/3\>4.6\>0.307\>--0.401\\
vacuum\>0\>--1\>--0.815\>0.662\>0.552\\
textures\>1\>--2/3\>1\>0.964\>0.824\\
``curvature''\>2\>--1/3\>--0.185\>2.294\>1.368\\
matter\>3\>0\>0\>\>2.677\\
radiation\>4\>+1/3\\
\\
The c,f,g coefficients have been evaluated at $\Omega_3$ = 0.27.
\end{tabbing}

\section{Conclusions}
Our primary conclusion is that introducing $\Omega_1$ or $\Omega_{-1}$ does not change WMAP values of H$_0$ or $\Omega_m$ .
It is easy to show that this conclusion extends to phantom energy generally for w $<$ --1 with $\Sigma a^{-n}\Omega_n~+~a^{-x}\Omega_x~
= ~h^2$ and x $<$ 0.
Using the supernova data alone, it is not possible to determine all the $\Omega$'s because of degeneracies.  But in combination with CMB
data, the degeneracies are broken.
Second, we find that $\Omega_1$ $<$ 0.2 and $\Omega_{-1}$ $<$ 0.1 with 95\% confidence.
Stricter limits will follow from dark energy program.
Third, $\Omega_2 ~\approx~ -0.2\Omega_1$ for $\Omega_1~ <<$ 1. If $\Omega_2~ = ~\epsilon$ (say, 
10$^{-6}$) due to inflation, $\Omega_1~<~0.018/(f_1-f_0)~=~0.06$.

\acknowledgements
I acknowledge very helpful discussions with Brian Schmidt and Chris Blake. Thanks go to Karl Glazebrook, Lucas Macri, and Paul Schechter for reading a draft and
to Ned Wright for compiling the observational data and making them available on his web page. This research is part of the Dark Universe scientific program of CAASTRO http://caastro.org
and supported by ARC.

\pagebreak

\pagebreak

\begin{figure}[h]


\begin{center}

\includegraphics[clip, width=\textwidth]{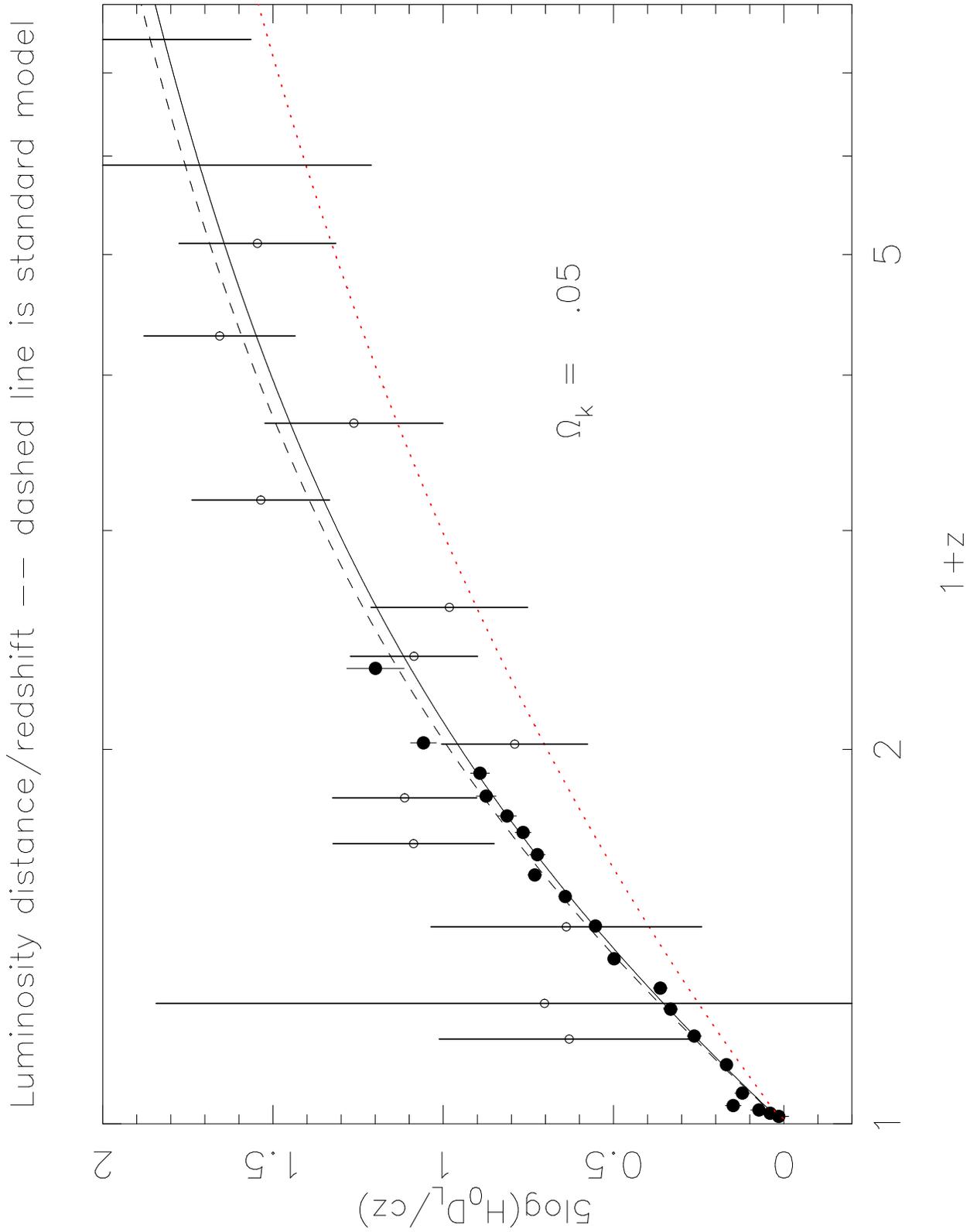}

\end{center}


\caption{ Curvature (solid curve for $\Omega_2$ = 0.05). The solid symbols are supernovae; the open symbols are GRBs. The standard model is the dashed curve. The red line dotted line shows that 0.5 violates the data.}

\end{figure}

\begin{figure}[h]



\includegraphics[clip,width=\textwidth]{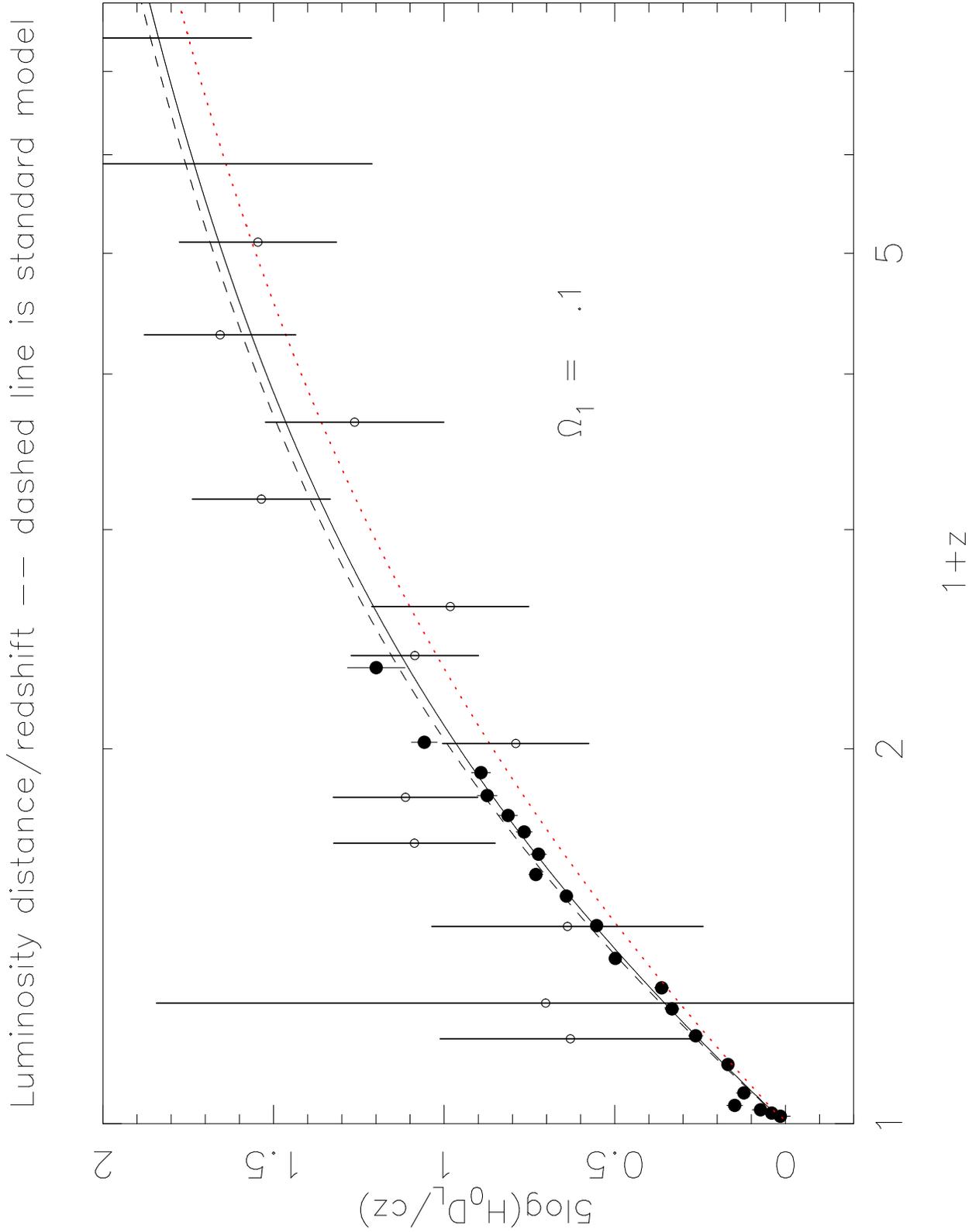}



\caption{Textures (solid curve for $\Omega_1$ = 0.1).  The standard model is the dashed curve. The red dotted line shows that 0.5 violates the data.}

\end{figure}

\begin{figure}[h]



\includegraphics[clip,width=\textwidth]{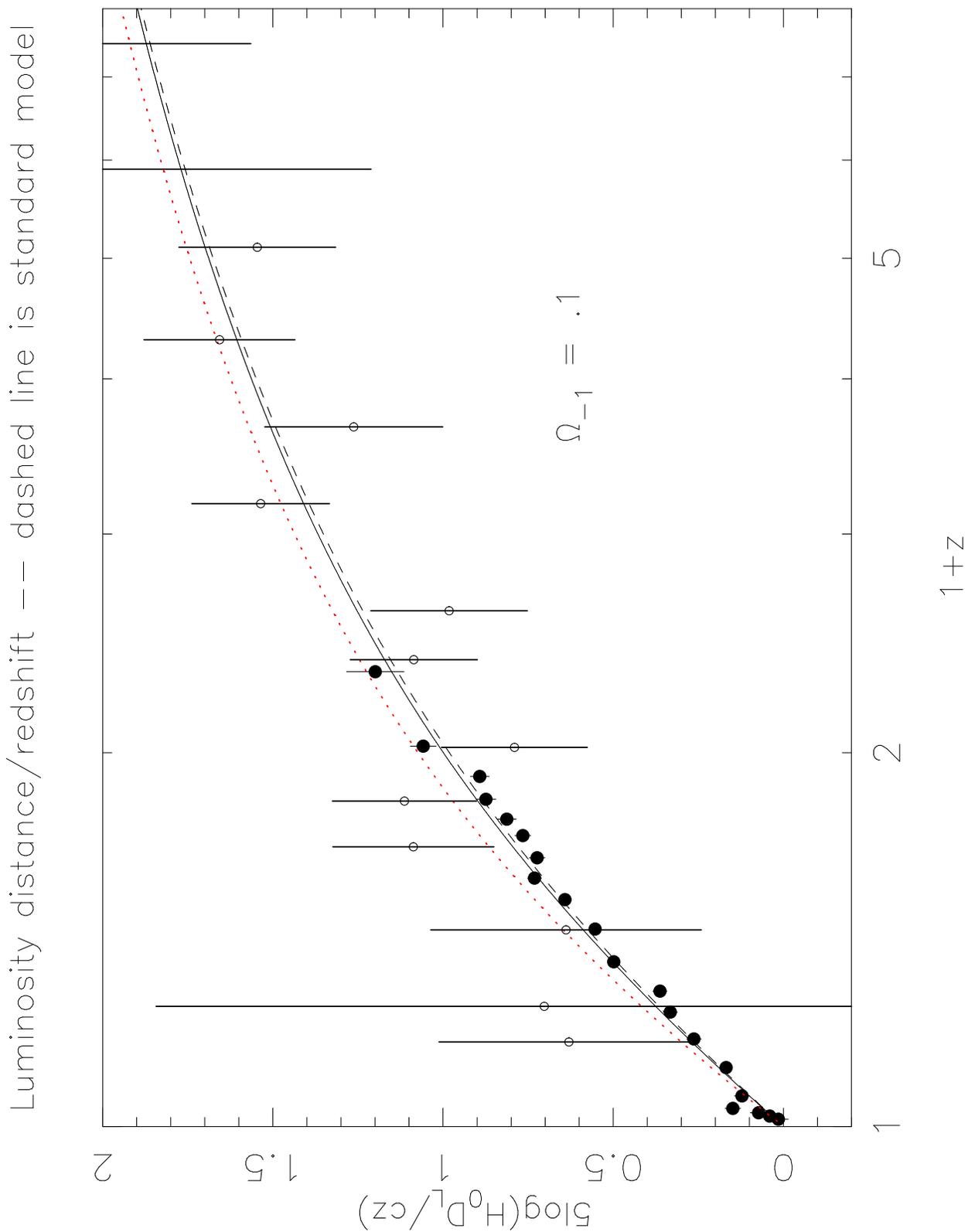}



\caption{Phantom power (solid curve for $\Omega_{-1}$ = 0.1).  The standard model is the dashed curve. The red dotted line shows that 0.5 violates the data.}

\end{figure}
\begin{figure}[h]



\includegraphics[clip,angle=-90,width=\textwidth]{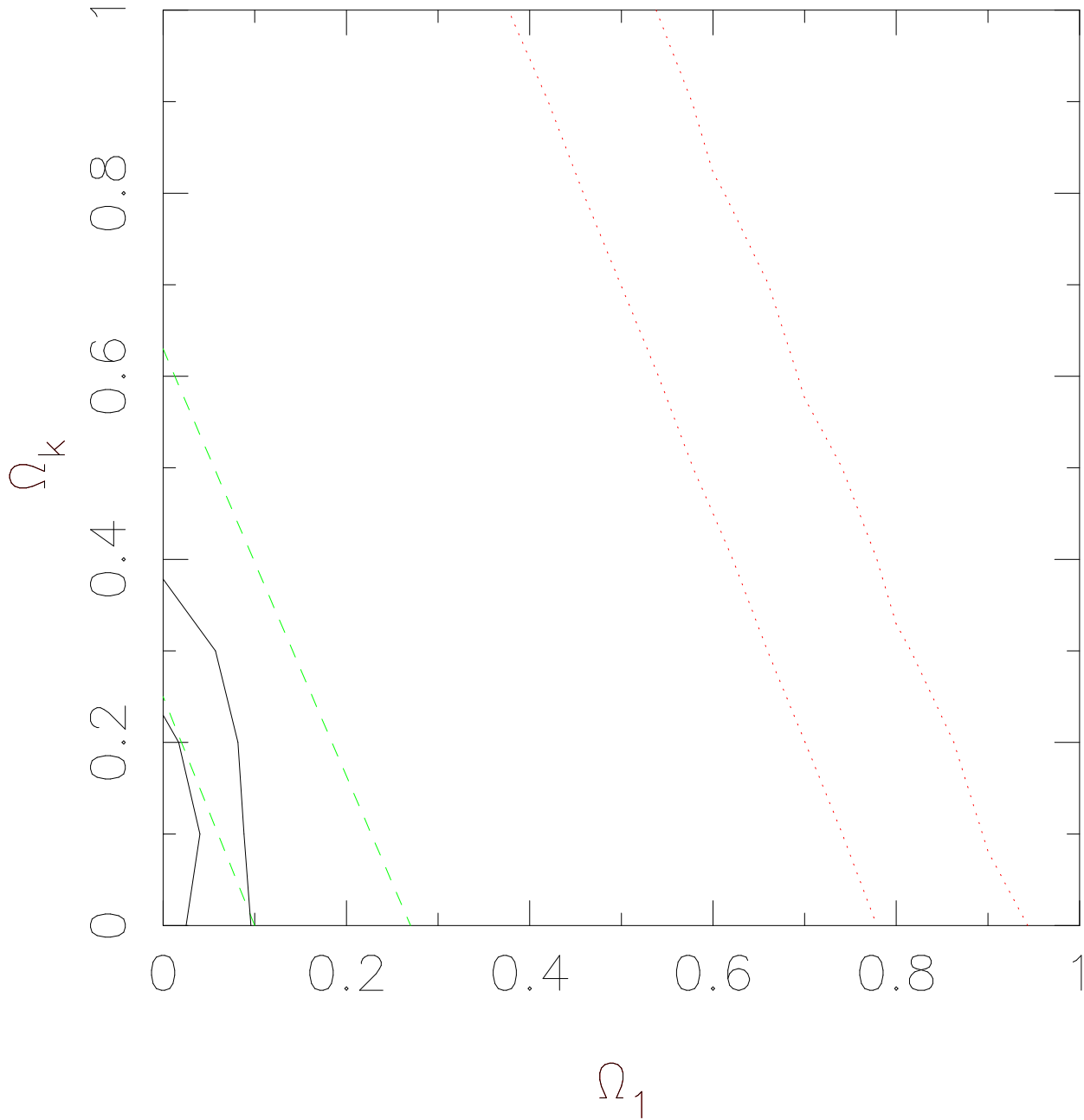}



\caption{Probability contours (1$\sigma$ and 2$\sigma$) for $\Omega_1$
and $\Omega_2$ given the WMAP7 CMB data and the supernova data in Figure 1. The SN contours are marked in red dots.
The green dashed lines show the probability contours if a prior is added to the supernova constraint, namely
$\Omega_3$ = 0.273 $\pm$ 0.025 (Eisenstein \etal 2005) from SDSS. Constraints from Baryon Acoustic Oscillation
experiments such as WiggleZ look like supernova constraints in this diagram, but are beyond the scope of the
present paper.}

\end{figure}
\begin{figure}[h]



\includegraphics[clip,angle=-90,width=\textwidth]{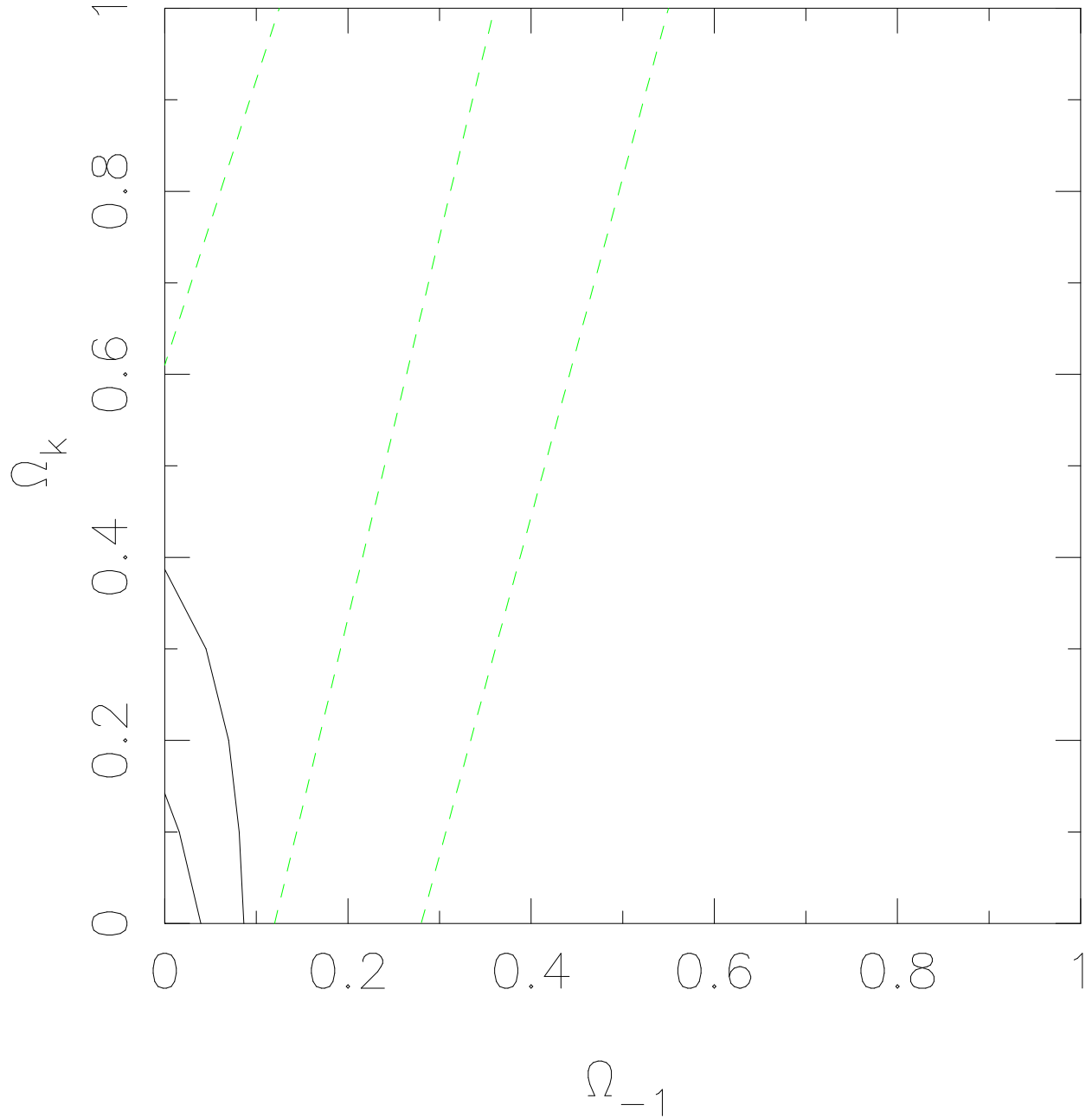}



\caption{Probability contours (1$\sigma$ and 2$\sigma$) for $\Omega_{-1}$
and $\Omega_2$ given the WMAP7 CMB data and the supernova data in Figure 1. 
The green dashed lines show the probability contours if a prior is added to the supernova constraint, namely
$\Omega_3$ = 0.273 $\pm$ 0.025 (Eisenstein \etal 2005) from SDSS.}

\end{figure}

\begin{thebibliography}{} 


\bibitem[Blake \etal (2011)]{B11}Blake, C. \etal 2011, astro-ph 1104.2948
\bibitem[Conley \etal (2011)]{F01}Conley, A. \etal 2011, \apjs, 192, 1
\bibitem[Eisenstein \etal (2005)]{E05}Eisenstein, D. \etal 2005, \apj, 633, 560
\bibitem[Frieman \etal (2005)]{F05}Frieman, J. and The Dark Energy Survey 2005, BAAS, 36, 1462
\bibitem[Garnavich \etal (1998)]{G98}Garnavich, P. \etal 1998, \apj, 509, 74

\bibitem[Gooding \etal (1992)]{G92}Gooding, A. \etal  1992, \apj, 393, 42
\bibitem[Komatsu \etal (2009)]{K09}Komatsu, S. \etal 2009, \apjs, 180, 330
\bibitem[Komatsu \etal (2011)]{K11}Komatsu, S. \etal 2011, \apjs, 192, 18
\bibitem[Peacock \etal (2001)]{P01}Peacock, J. \etal 2001, Nature, 410, 169
\bibitem[Perlmutter \etal (1999)]{P99}Perlmutter, S. \etal 1999, \apj, 517, 565
\bibitem[Schaefer (2006)]{R11}Schaefer, B. 2007, \apj, 660, 16
\bibitem[Turner (1999)]{T99}Turner, M. 1999, The Third Stromlo Symposium: The Galactic Halo, ASP Conference Series, 165, 431

\end{thebibliography}
\end{document}